\documentclass[%
 reprint,
 amsmath,amssymb,
 aps,
floatfix,
]{revtex4-1}

\usepackage{graphicx,subfigure,lipsum}
\usepackage{dcolumn}
\usepackage{color}
\usepackage{bm}
\usepackage{graphicx}
\usepackage{algorithm}
\usepackage{algorithmic}

\begin{document}

\preprint{APS/123-QED}

\title{Dodge and survive: modeling the predatory nature of dodgeball}

\author{Perrin E. Ruth}
\email{Perrin.Ruth@colorado.edu}
\author{Juan G. Restrepo}%
 \email{juanga@colorado.edu}
\affiliation{%
 Department of Applied Mathematics, University of Colorado at Boulder, Boulder, CO 80309, USA
}%


\
\date{\today}

\begin{abstract}
The analysis of games and sports as complex systems can give insights into the dynamics of human competition, and has been proven useful in soccer, basketball, and other professional sports. In this paper we present a model for dodgeball, a popular sport in US schools, and analyze it using an ordinary differential equation (ODE) compartmental model and stochastic agent-based game simulations. The ODE model reveals a rich landscape with different game dynamics occurring depending on the strategies used by the teams, which can in some cases be mapped to scenarios in competitive species models. Stochastic agent-based game simulations confirm and complement the predictions of the deterministic ODE models. In some scenarios, game victory can be interpreted as a noise-driven escape from the basin of attraction of a stable fixed point, resulting in extremely long games when the number of players is large. Using the ODE and agent-based models, we construct a strategy to increase the probability of winning.
\end{abstract}

\maketitle

\section{Introduction}

Games and sports are emerging as a rich testbed to study the dynamics of competition in a controlled environment. Examples include the analysis of passing networks \cite{Buletal2018,mchale2018identifying} and entropy \cite{martinez2020spatial} in soccer games (see also \cite{rein2016big} for a discussion on data-driven tactical approaches), scoring dynamics \cite{MerCla2014, clauset2015safe,kiley2016game} and play-by-play modeling \cite{vravcar2016modeling,wang2019tac} in professional sports such as hockey, basketball, football, and table tennis, penalty kicks in soccer games \cite{Pal03}, and serves in tennis matches \cite{WalWoo01}. 
Here we explore the dynamics of \textit{dodgeball}, where the number of players playing different roles changes dynamically and ultimately determines the outcome of the game.    While modeling dodgeball might seem like a very specific task, it is a relatively clean and well-defined system where the ability of mean-field techniques \cite{lasry2007mean,bensoussan2013mean} to describe  human competition can be put to the test. In addition, it complements ongoing efforts to quantify and model dynamics in sports and games \cite{Buletal2018,mchale2018identifying, martinez2020spatial, MerCla2014,rein2016big, clauset2015safe,kiley2016game, vravcar2016modeling,wang2019tac,Pal03,WalWoo01}.


In this paper we present and analyze  a mathematical model of dodgeball based on both agent-based stochastic game simulations and an ordinary differential equation (ODE) based compartmental model. By analyzing the stability of fixed points of the ODE system, we find that different game dynamics can occur depending on the teams' strategies: one of the teams achieves a quick victory, either team can achieve a victory depending on initial conditions, or the game evolves into a stalemate. For the simplest strategy choice, these regimes can be interpreted in the context of a competitive Lotka-Volterra model. Numerical simulations of games based on stochastic behavior of individual players reveal that the stalemate regime corresponds to extremely long games with large fluctuations. These long games can be interpreted as a noise-driven escape from the basin of attraction of the stable stalemate fixed point, and are commonly observed in dodgeball games (see Fig.~\ref{fig:game1}). Using both the stochastic and ODE models, we develop a greedy strategy and demonstrate it using stochastic simulations.

The structure for the paper is as follows. In Section \ref{sec:game} we describe the rules of the game we will analyze. In Section \ref{sec:dyn} we present and analyze a compartment-based model of  dodgeball. In Section \ref{sec:stoch} we present stochastic numerical simulations of dodgeball games and compare these with the predictions of the compartmental model. We then discuss the notion of strategy in the context of this stochastic model. Finally, we present our conclusions in Sec.~\ref{sec:conclusions}.

\section{Description of Dodgeball}
\label{sec:game}

\begin{figure}[b]
   \centering
    \includegraphics[width=0.85\columnwidth]{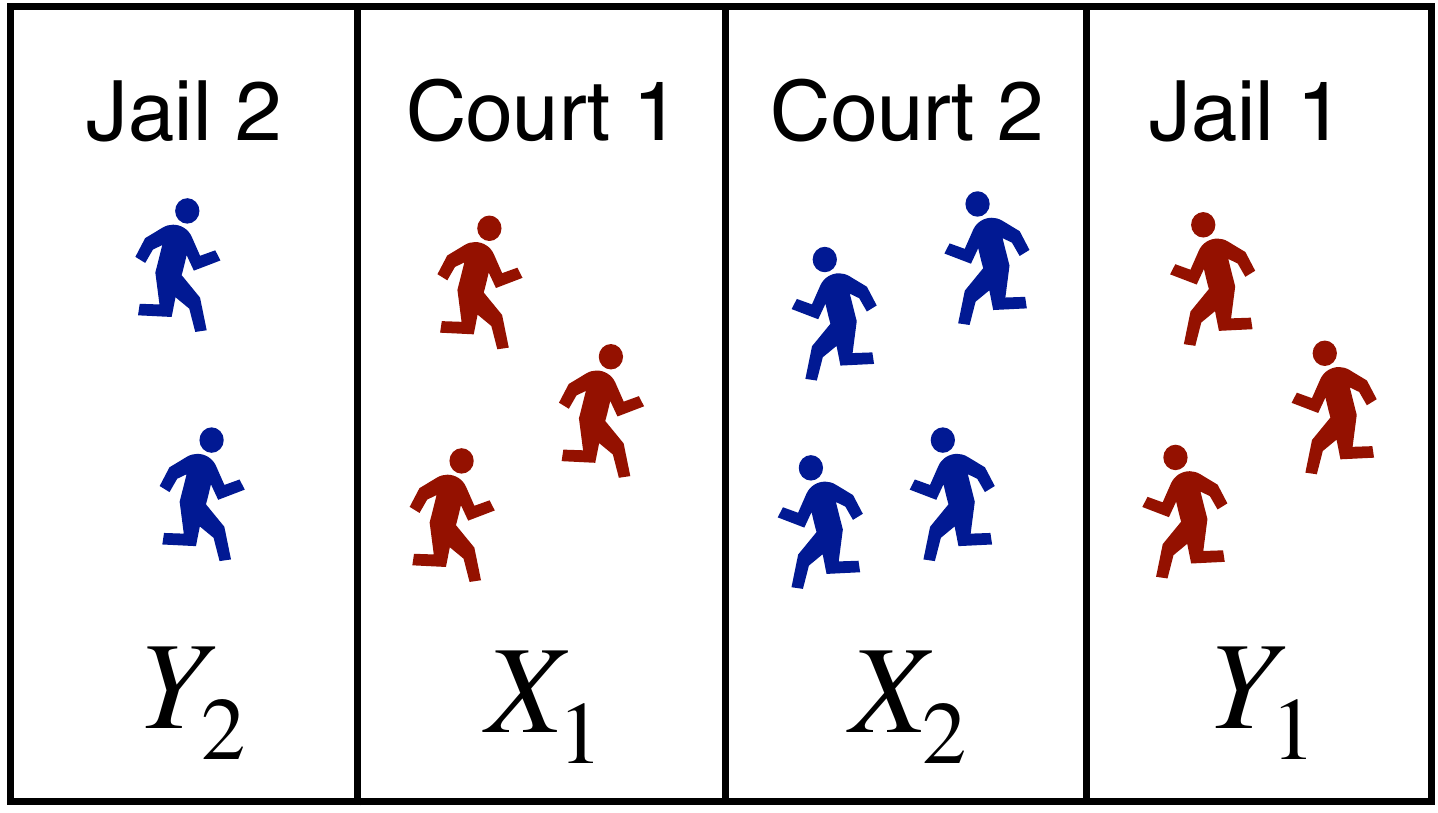}
    \caption{(a) Setup of dodgeball court. Players in team $i$ make transitions between Court $i$ and Jail $i$, and Team $i$ loses when there are no players in court $i$.}
    \label{fig:game0}
\end{figure}

In this paper we consider the following variant played often in elementary schools in the US (sometimes called {\it prison dodgeball}). Two teams (Team 1 and Team 2) of $N$ players each initially occupy two zones adjacent to each other, which we will refer to as Court 1 and Court 2 (see Fig.~\ref{fig:game0}). Players in a Court can throw balls at players of the opposite team in the other Court. If a player in a Court is hit by such a ball, they move to their respective team's \textit{Jail}, an area behind the opposite team's Court. A player in a Court may also throw a ball to a player of their own Team in their Jail, and if the ball is caught, the catching player returns to their Team's Court.  These processes are illustrated schematically in Fig.~\ref{fig:comic}. We denote the number of players on Team $i$ that are in Court $i$ and Jail $i$ by $X_i$ and $Y_i$, respectively. Team $i$ loses when $X_i=0$. For simplicity, we assume there are always available balls and neglect the possibility that a player catches a ball thrown at them by an enemy player. 

\begin{figure}[t]
    \begin{subfigure}
    \centering
    \includegraphics[width=\linewidth]{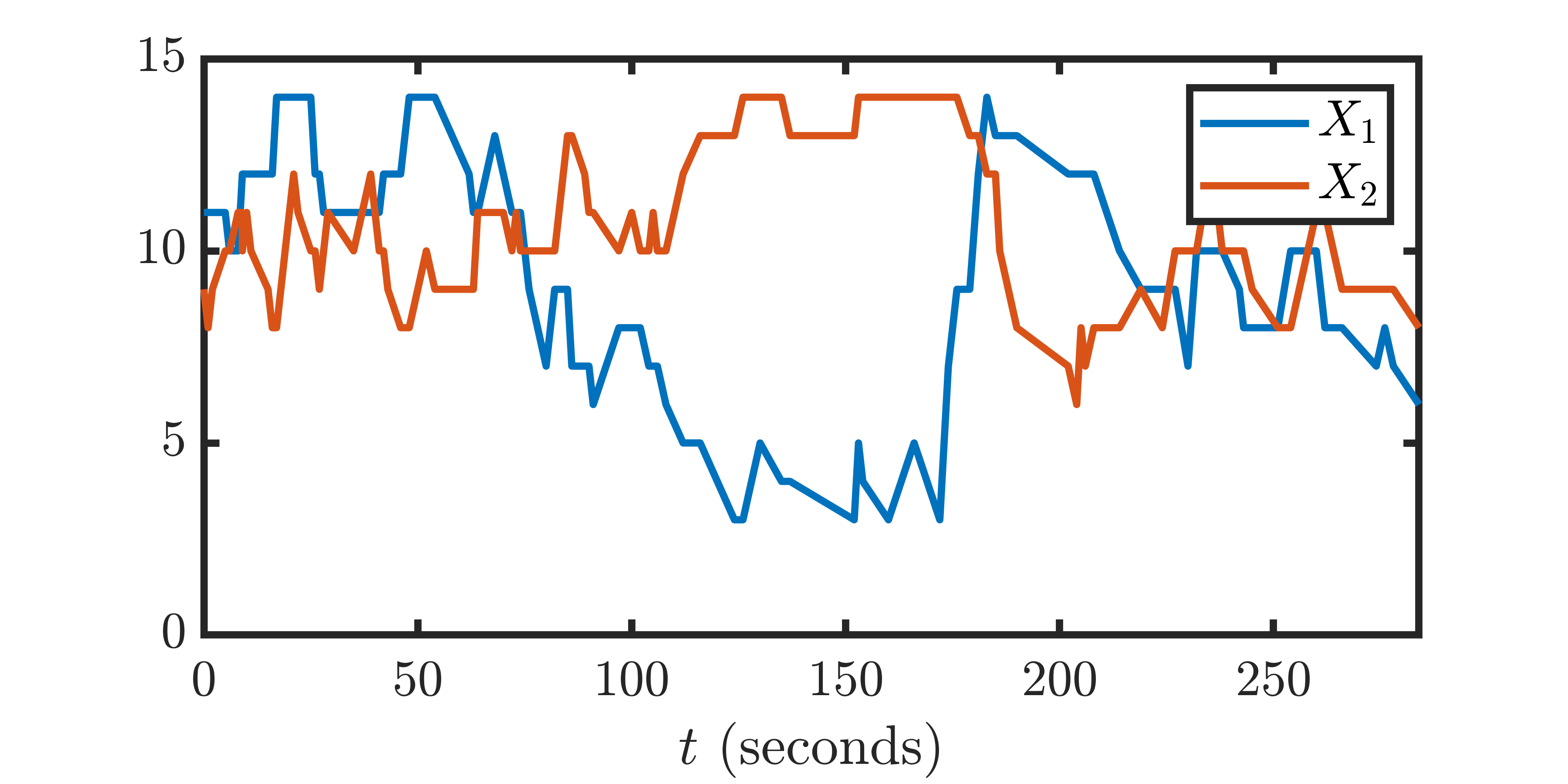}
    \end{subfigure}
    \begin{subfigure}
    \centering
    \includegraphics[width=\linewidth]{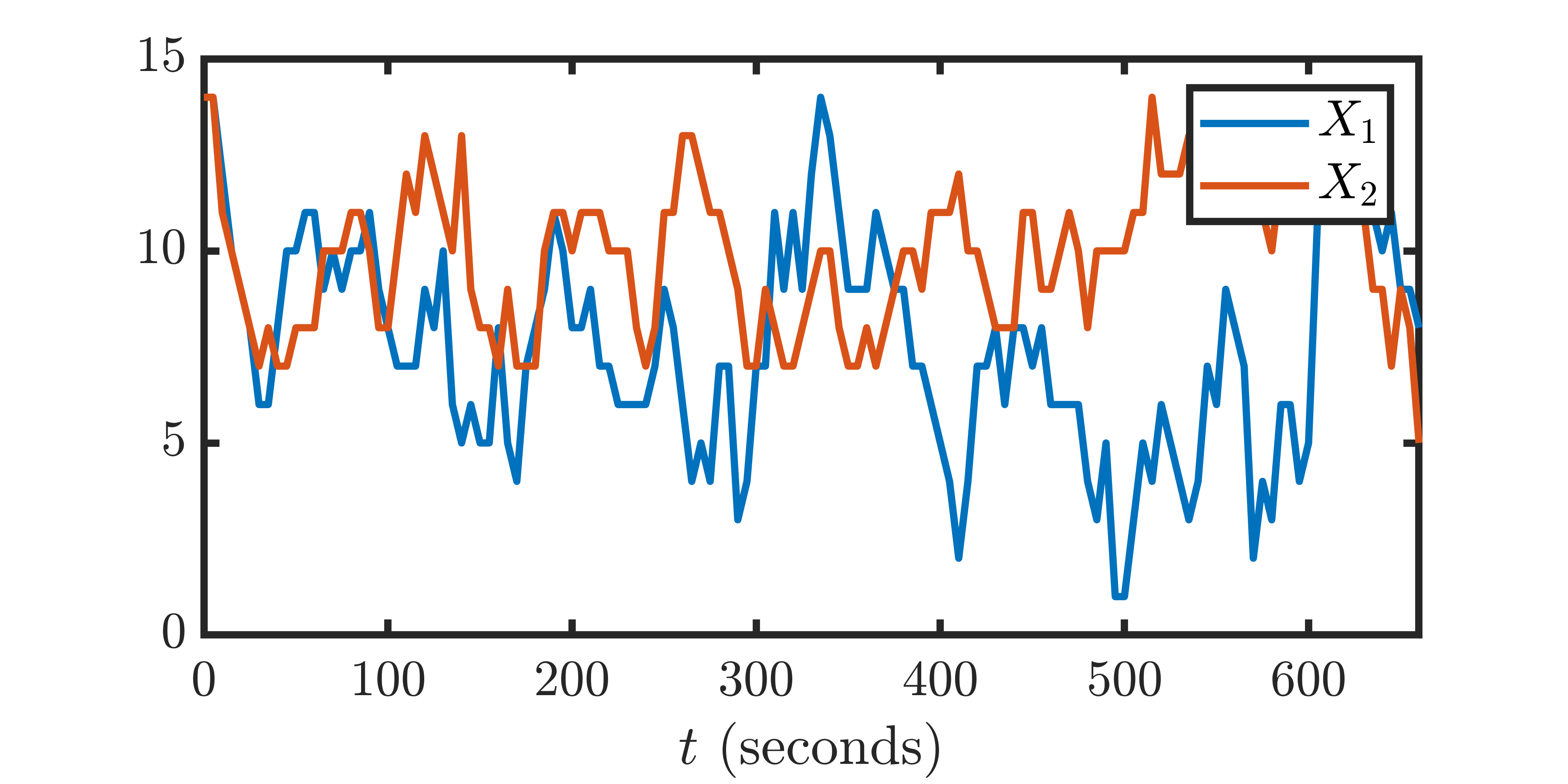}
    \end{subfigure}
    \caption{Evolution of two fifth-grade dodgeball games played in Eisenhower Elementary in Boulder, Colorado, USA. The number of players in Courts $1$ and $2$, $X_1$ and $X_2$, fluctuate for a long time without any team gaining a decisive advantage. The games were eventually stopped and a winner decided on the spot.}
    \label{fig:game1}
\end{figure}

In practice, games often last a long time without any of the Teams managing to send all the enemy players to Jail. Because of this, such games are stopped at a predetermined time and the winner is decided based on other factors (e.g., which Team has more players on their Court). An example of this is in Figure~\ref{fig:game1}, which shows the numbers of players in Courts $1$ and $2$, $X_1$ and $X_2$, during two fifth-grade dodgeball games  in Eisenhower Elementary in Boulder, Colorado. The values of $X_1$ and $X_2$ seem to fluctuate without any team obtaining decisive advantage. The games continued after the time interval shown and were eventually stopped. Our subsequent model and analysis suggests that this stalemate behavior is the result  of underlying dynamics that has a stable fixed point about which $X_1$ and $X_2$ fluctuate.

\section{Rate Equation description of game dynamics}
\label{sec:dyn}

\begin{figure}[b]
    \centering
    \includegraphics[width=\linewidth]{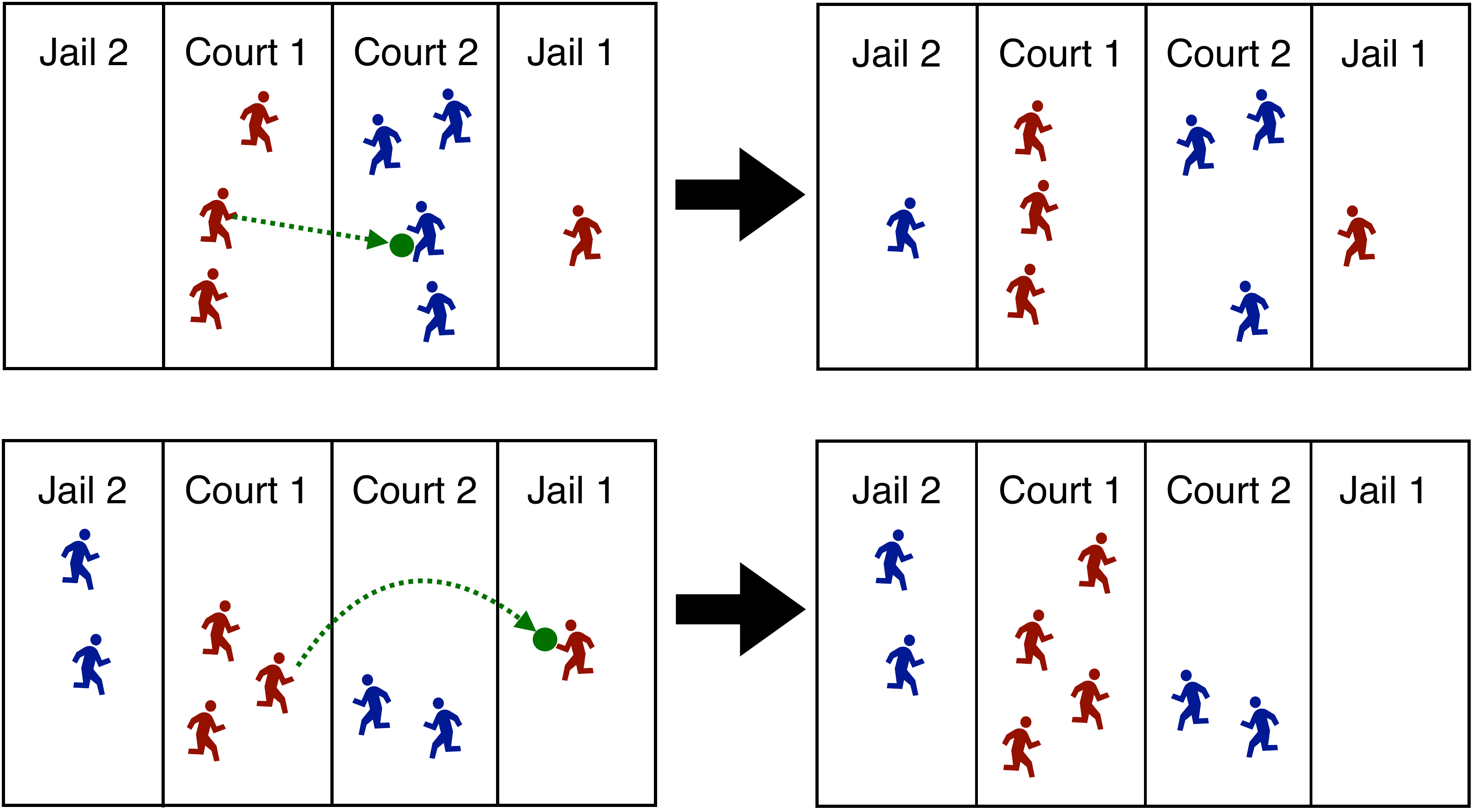}
    \caption{(Top) A player in a Court can be sent to Jail when hit by a ball from a player in the opposing Court. (Bottom) A player can be saved from Jail when catching a ball thrown by a player from their Court.}
    \label{fig:comic}
\end{figure}

We begin our description of the game dynamics by adopting a continuum formulation where the number of players in Courts $1$ and $2$ are approximated by continuous variables. These variables evolve following rate equations obtained from the rates at which the processes described in the previous section and illustrated in Fig.~\ref{fig:comic} occur. Since the number of players in a dodgeball game is not too large (typically less than $50$), and the game is decided when the number of players in a court drops to zero, one might question the validity of a continuum description. However, as we will see in Sec \ref{sec:stoch}, stochastic simulations with few players show that the rate equations give useful insights about the dynamics of simulated games with a finite number off players. 

To construct the rate equations, we define $\lambda$ as the mean throw rate of the players. Consequently, team $i$ throws balls at a rate of $\lambda X_i$. We also define $F_i(X_1,X_2)$ as the fraction of balls that team $i$ throws that are directed at enemy players, $p_e(X)$ as the probability that a ball thrown at $X$ opposing players hits one of them, and $p_j(Y)$ as the probability that a ball thrown at $Y$ players in jail is caught. Combining these processes and using $Y_i = N-X_i$ we get the Dodgeball Equations:
\begin{align}
    \begin{split}
    \dot{X}_1 &= \lambda X_1[1-F_1(X_1,X_2)]p_j(N_1-X_1)\\
    &- \lambda X_2F_2(X_1,X_2)p_e(X_1),  \label{eq:genModel_a}
    \end{split}\\
    \begin{split}
    \dot{X}_2 &= \lambda X_2[1-F_2(X_1,X_2)]p_j(N_2-X_2)\\
    &-\lambda X_1F_1(X_1,X_2)p_e(X_2).
    \label{eq:genModel_b}
    \end{split}
\end{align}
Note that, given the initial conditions $X_i(0)=N$, $X_i(t) \in [0,N]$ for all $t\ge0$.
 For simplicity, we assume the functions $p_j$ and $p_e$ to be linear, $p_j(Y)=k_j Y$ and $p_e(X)=k_e X$. Defining the normalized number of players $x_i = X_i/N\in [0,1]$ and the dimensionless time $\tau = \lambda N k_j t$, we get the simplified Dodgeball Equations:
\begin{align}
    \frac{dx_1}{d\tau} &= x_1(1-x_1) [1-{f_1}(x_1,x_2)] -  {c} x_1 x_2 {f_2}(x_1,x_2),
    \label{eq:simplified_a}\\
    \frac{dx_2}{d\tau} &= x_2(1-x_2) [1-{f_2} (x_1,x_2)]-{c} x_1 x_2 {f_1}(x_1,x_2),
    \label{eq:simplified_b}
\end{align}
where ${f_i}(x_1,x_2) = F_i(N x_1,N x_2)$ and ${c} = k_e/k_j>0$ is the effectiveness of throwing a ball at an enemy relative to throwing a ball at jail.

\begin{table}[t]
\setlength\tabcolsep{0pt} 
\smallskip 
\begin{tabular*}{\columnwidth}{@{\extracolsep{\fill}}|c|c|}
  \hline
  Symbol & Meaning\\
  \hline
  $a_i$ & Probability that a player in Team i tries to hit an\\ &opponent instead of saving a teammate from jail\\
  \hline
  $x_i$ & Fraction of players in Team i in Court i\\   
  \hline
  $c$ & Probability of hitting/probability of saving\\
  \hline
\end{tabular*}
\caption{Notation used in the dodgeball model Equations~\eqref{eq:fixed_strat_a}-\eqref{eq:fixed_strat_b}.} 
\label{tab:freq}
\end{table}


\subsection{Example: fixed strategy}

As an illustrative example we will focus on the case when the strategy for both teams is fixed over the course of the game, ${f_i}(x_1,x_2)=a_i\in(0,1)$. We will consider state-dependent choices for $f_i$ (i.e., strategies)  in Sec.~\ref{sec:stoch}. Inserting $f_i(x_1,x_2) = a_i$ into Equations (\ref{eq:simplified_a})-(\ref{eq:simplified_b}) gives
\begin{align}
    \frac{dx_1}{d\tau} &= x_1(1-x_1) (1-a_1) - {c} x_1 x_2 a_2
    \label{eq:fixed_strat_a},\\
    \frac{dx_2}{d\tau} &= x_2(1-x_2) (1-a_2)-{c} x_1 x_2 a_1
    \label{eq:fixed_strat_b},
\end{align}
which is a 2-species competitive Lotka-Volterra system \cite{gotelli2001primer}. In this case, we can use known results about this system to understand the possible game scenarios. Specifically, at $\tau = 0$ the system starts at $(x_1,x_2) = (1,1)$. For $\tau > 0$, the solution converges towards one of the stable fixed points of (\ref{eq:fixed_strat_a})-(\ref{eq:fixed_strat_b}) in the invariant square $[0,1]\times[0,1]$, which  are $(0,0)$, $(0,1)$, $(1,0)$, and the solutions $(x_1^*,x_2^*)$ of the linear system
\begin{align}
    0 &= (1-x_1) (1-a_1) - {c} x_2 a_2 \label{eq:fix_line_a},\\
    0 &= (1-x_2) (1-a_2) - {c} x_1 a_1\label{eq:fix_line_b}.
\end{align}
If $a_1 a_2 c^2 \neq (1-a_1)(1-a_2)$ there is a unique solution to these equations, the fixed point 
\begin{align}
    x_1^* &= \frac{(1-a_2)[a_2 {c}-(1-a_1)]} {a_1 a_2 {c}^2-(1-a_1)(1-a_2)},\\
    x_2^* &= \frac{(1-a_1)[a_1 {c} - (1-a_2)]}{a_1 a_2 {c}^2-(1-a_1)(1-a_2)}.
    \label{xstar}
\end{align}
The degenerate case where $a_1 a_2 c^2 = (1-a_1)(1-a_2)$ gives a continuum of fixed points described by
\begin{equation}
    x_1^*+x_2^* = 1, \label{eq:fixed_point_line2}
\end{equation}
when $a_1 = (1-a_2)/c$ and $a_2 = (1-a_1)/c$,  and no solution otherwise. 
\begin{figure}[t]
\centering
\includegraphics[width=\linewidth]{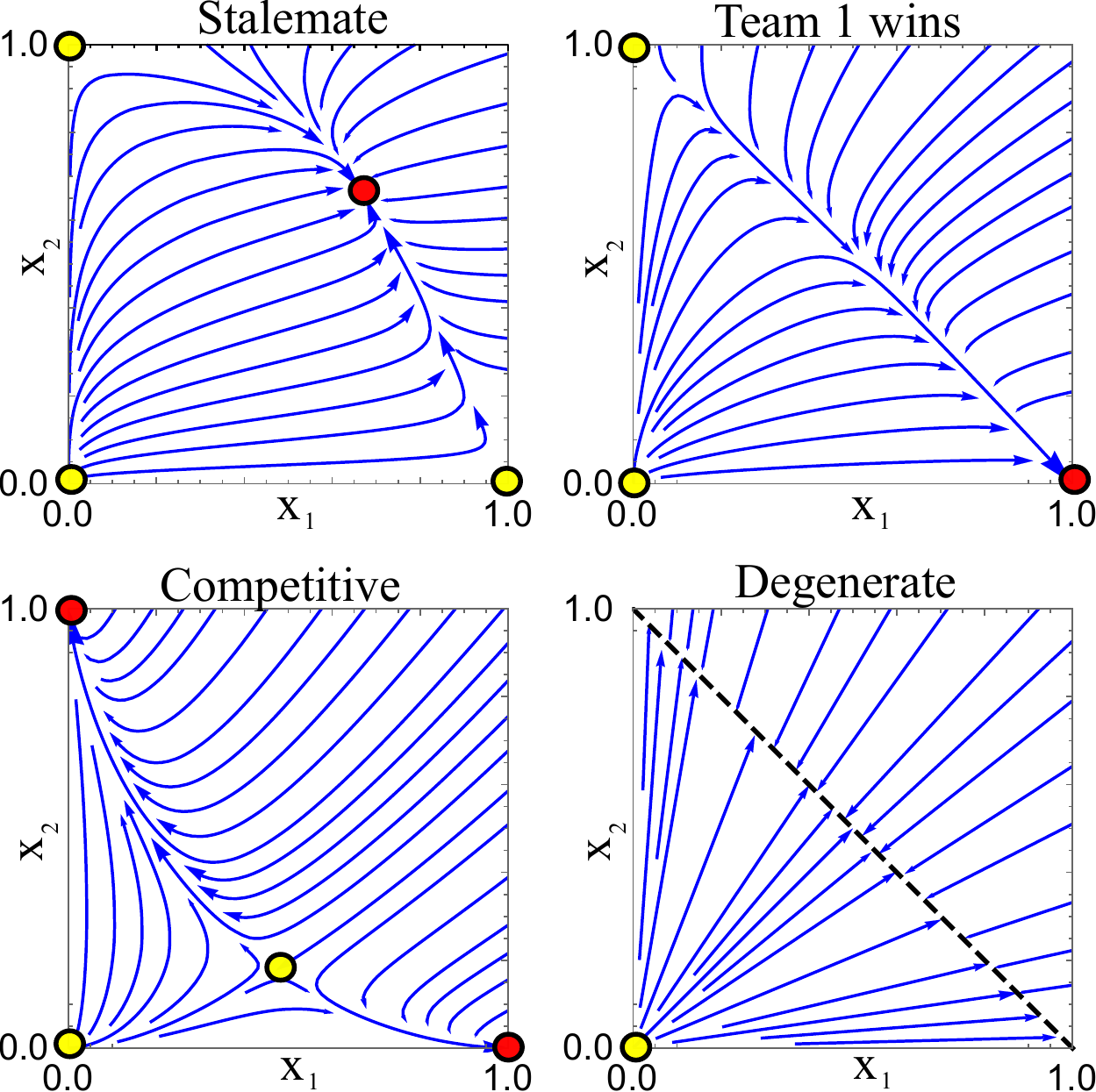}
\caption{Stream plots of Equations (\ref{eq:fixed_strat_a}-\ref{eq:fixed_strat_b}) with ${c} = 0.5$ and various values of $a_1$ and $a_2$. (Top left) {\it Stalemate}: for $a_1 = 1/4$, $a_2 = 3/4$, both $(0,1)$ and $(1,0)$ are unstable and $(x_1^*,x_2^*)$ is stable. (Top right) {\it Team 1 wins}: for for $a_1 = 9/16$, $a_2 = 3/4$, $(1,0)$ is a stable fixed point while $(0,1)$ is unstable, giving Team $1$ the advantage; note that in this case $(x_1^*,x_2^*)\notin [0,1]^2$. (Bottom left) {\it Competitive}: for $a_1 = 7/8$, $a_2 = 3/4$, both $(0,1)$ and $(1,0)$ are stable fixed points, and the winner is determined by the initial conditions. (Bottom right) {\it Degenerate}: For the special case $a_1=a_2=(1+c)^{-1}$, every point on the line $x_1+x_2=1$ is a fixed point.}
\label{fig:streams}
\end{figure}

The fixed point $(0,0)$  corresponds to both teams running out of players, 
the fixed points $(1,0)$ and $(0,1)$ correspond to Team $1$ and Team $2$ winning, respectively, and the fixed point $(x_1^*,x_2^*)$, when it is stable and in $(0,1)^2$, corresponds to a stalemate situation where the number of players in each court remains constant in time. 
By analyzing the linear stability of the fixed points (see, e.g., \cite{gotelli2001primer}), one finds  that the game dynamics can be classified in the following cases:
\begin{itemize}
\item {\it Stalemate.} This occurs when $(0,1)$, $(1,0)$ are both unstable and  $(x_1^*,x_2^*)$ is in $[0,1]^2$ and is stable, which occurs when $a_1 < (1-a_2)/c$ and $a_2 < (1-a_1)/c$. In this scenario, the solution settles in the fixed point $(x_1^*,x_2^*)$ and no Team wins in the deterministic version of the game. The flow corresponding to this case is shown in Fig.~\ref{fig:streams} (top left). This scenario is analogous to the ``Stable coexistence'' of species in the Lotka-Volterra model.
\item {\it Competitive.} This occurs when $(0,1)$, $(1,0)$ are stable and the fixed point $(x_1^*,x_2^*)$ is in $[0,1]^2$ and is unstable, which occurs when $a_1 > (1-a_2)/c$ and $a_2 > (1-a_1)/c$. The stable manifold of $(x_1^*,x_2^*)$ acts as a separatrix for the basins of attraction of the fixed points that correspond to victories for Team 1 and Team 2. See Fig.~\ref{fig:streams} (bottom left). This scenario is analogous to the ``Unstable coexistence'' of species in the Lotka-Volterra model.
\item {\it Team 1 wins.} This occurs when $(0,1)$ is unstable and $(1,0)$ is stable, which occurs when $a_1 > (1-a_2)/c$ and $a_2 < (1-a_1)/c$. In this scenario, the solution converges towards a victory by Team 1. See Fig.~\ref{fig:streams} (top right). This scenario is analogous to the ``Competitive exclusion'' of species in the Lotka-Volterra model, in which one species is driven to extinction by the other.
\item {\it Team 2 wins.} This occurs when $(0,1)$ is stable and $(1,0)$ is unstable, and is analogous to the {\it Team 1 wins} case. In this scenario, the solution converges towards a victory by Team 2.
\item {\it Degenerate.} This occurs when there is a continuum of fixed points $x_1^*+x_2^* = 1$. In this scenario, the solution converges towards the line $x_1 + x_2 = 1$, and no winner is produced in the deterministic version of the game. See Fig.~\ref{fig:streams} (bottom right).
\end{itemize}

\begin{figure}[b]
    \includegraphics[width=\linewidth]{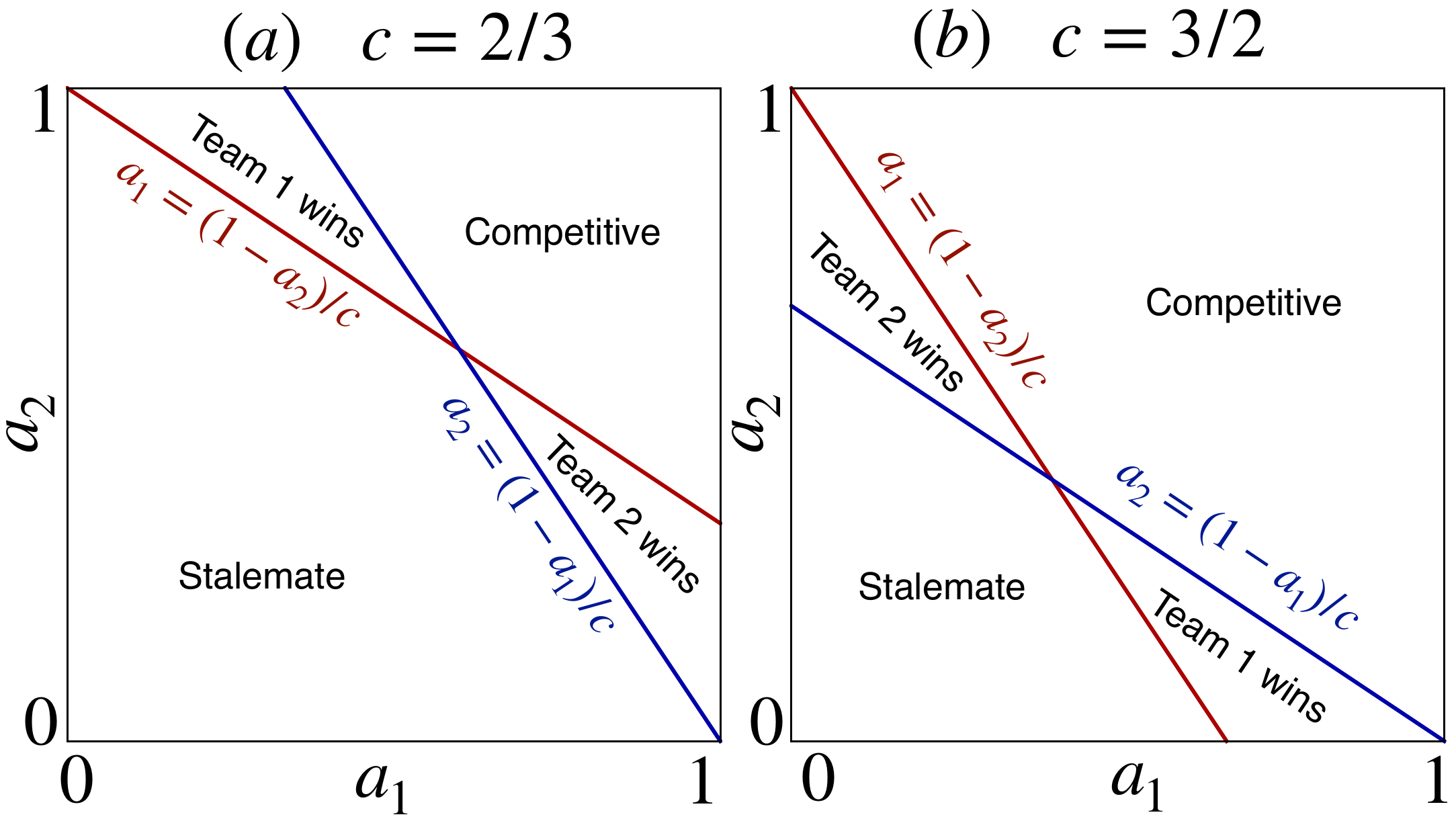}
    \caption{Deterministic game outcomes based on different strategies $(a_1,a_2)$ for  (a) $c<1$, (b) $c>1$.}
    \label{fig:fix_point guide}
\end{figure}

Figure \ref{fig:streams} illustrates these different game dynamics by showing the flow induced by Eqs.~(\ref{eq:fixed_strat_a})-(\ref{eq:fixed_strat_b}) in the region $0 \leq x_1 \leq 1$, $0\leq x_2 \leq 1$ for various parameter choices. Stable fixed points are shown as red circles, and unstable fixed points as yellow circles.

In Figure~\ref{fig:fix_point guide} we illustrate how the game outcome depends on the strategies used by both teams. The cases $c>1$ and $c<1$ are illustrated in  Figs.~\ref{fig:fix_point guide} (a) and (b), respectively. The strategy phase space $(a_1,a_2)$ is divided into four regions separated by the lines $a_1 = (1-a_2)/c$ and $a_2 = (1-a_1)/c$. When both teams preferentially save players of their own team from jail, instead of trying to hit players from the other team (i.e., both $a_1$ and $a_2$ are small), the game results in a stalemate (we reiterate that when stochasticity is included, this scenario corresponds to long games). When both teams preferentially hit players from the other team (i.e., bot $a_1$ and $a_2$ are close to $1$) a winner emerges quickly. When teams have opposite strategies, one of the teams can quickly win, depending on the value of $c$.


While the rate equation description provides interesting insights, it relies on the assumption of an infinite number of players. Because of this, some of its predictions are not reasonable for games with a finite number of players. For example, it predicts that the outcome of games is completely determined by parameters and initial conditions. In reality, games are determined by the aggregate behavior of a finite number of individual players, and chance can play an important role. In the next section we will model dodgeball games by considering the stochastic behavior of individual players, and we will find that the insights provided by the rate equations are useful to understand the stochastic dodgeball games.

\section{Stochastic Dodgeball Simulations}
\label{sec:stoch}

In this Section we present numerical simulations of dodgeball games using a stochastic agent-based model that corresponds to the simplified model used in Section~\ref{sec:dyn}. 
\begin{figure}[b]
    \centering
    \includegraphics[width=0.7\linewidth]{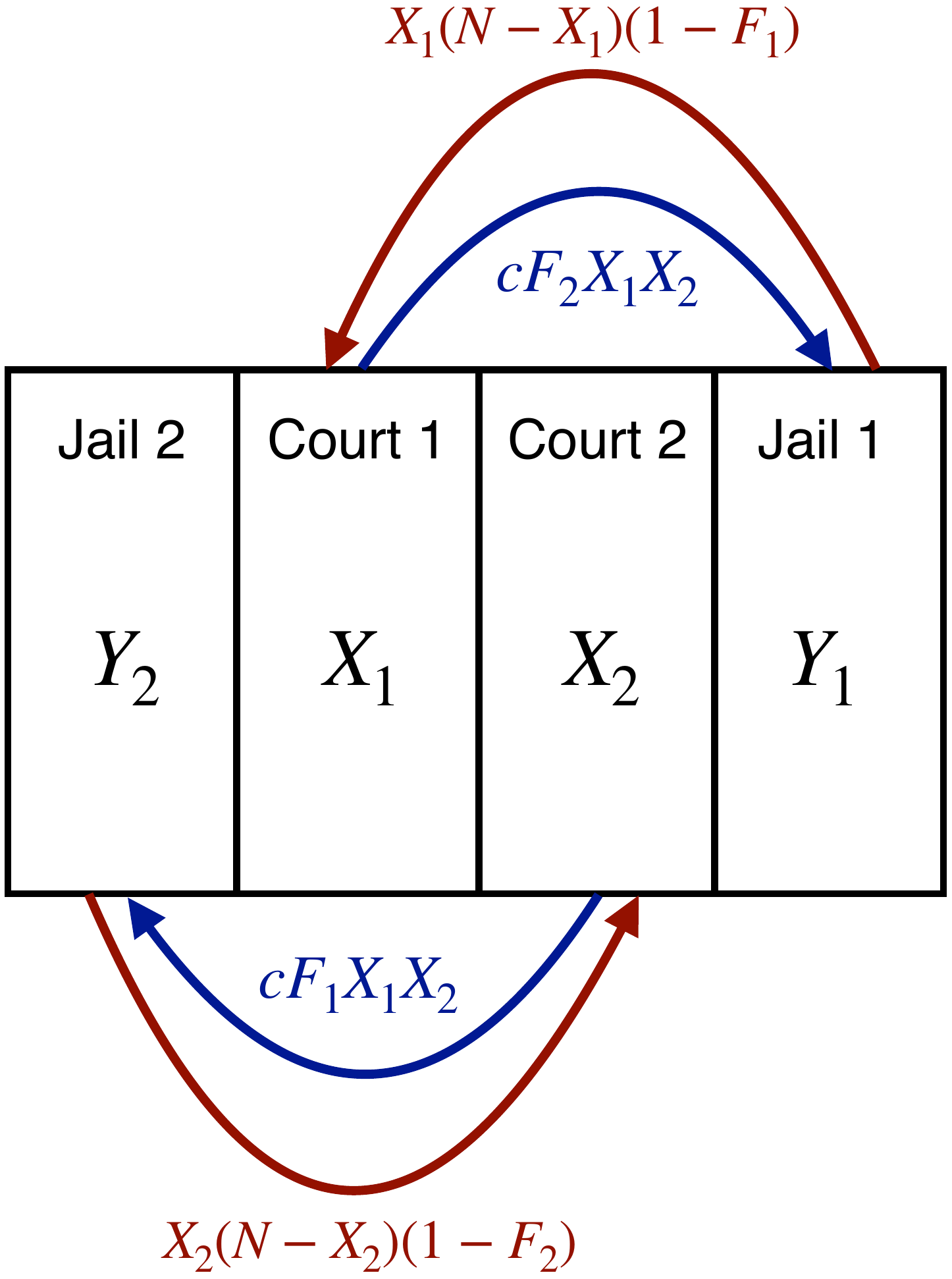}
    \caption{Stochastic dodgeball game. Players make transitions between the indicated compartments with the rates shown next to the arrows. The game ends when either $X_1 = 0$ or $X_2 = 0$.}
    \label{fig:cartoon}
\end{figure}

In the stochastic version of the game, each team starts with $N$ players in their respective court, $X_1(0) = X_2(0) = N$, and no players in Jail, $Y_1(0) = Y_2(0) = 0$. Players in Court $1$ make stochastic transitions to Jail $1$ at rate $\lambda X_2(t) F_2(X_1,X_2) k_e X_1$, and players in Jail $1$ make transitions to Court $1$ at rate $\lambda X_1 [1-F_1(X_1,X_2)] k_j (N-X_1)$, where, as in Sec.~\ref{sec:dyn}, $F_i(X_1,X_2)$ is the probability that a player in Court $i$ will throw a ball towards an enemy player in the opposite Court instead of trying to save a teammate from Jail, $k_e$ is the probability of hitting a single enemy player, and $k_j$ is the probability that a player in Jail catches a ball thrown at them. The rates of transition for players in Team 2 are obtained by permuting the indices $1$ and $2$. By using the dimensionless time $\tau = \lambda k_j t$,  the rates of transition per dimensionless time are  $c X_1 X_2 F_2(X_1,X_2)$ and $X_1 (N-X_1) [1-F_1(X_1,X_2)]$ for players to transition from Court 1 to Jail 1 and from Jail 1 to Court 1, respectively, where $c = k_e/k_j$. The compartmental model corresponding to this process is shown schematically in Fig.~\ref{fig:cartoon}. The code used for simulating the agent-based dodgeball model and finding the probability that a team wins can be found on the GitHub repository (\url{https://github.com/Dodgeball-code/Dodgeball}).

\begin{figure}[t]
    \centering
    \includegraphics[width=\linewidth]{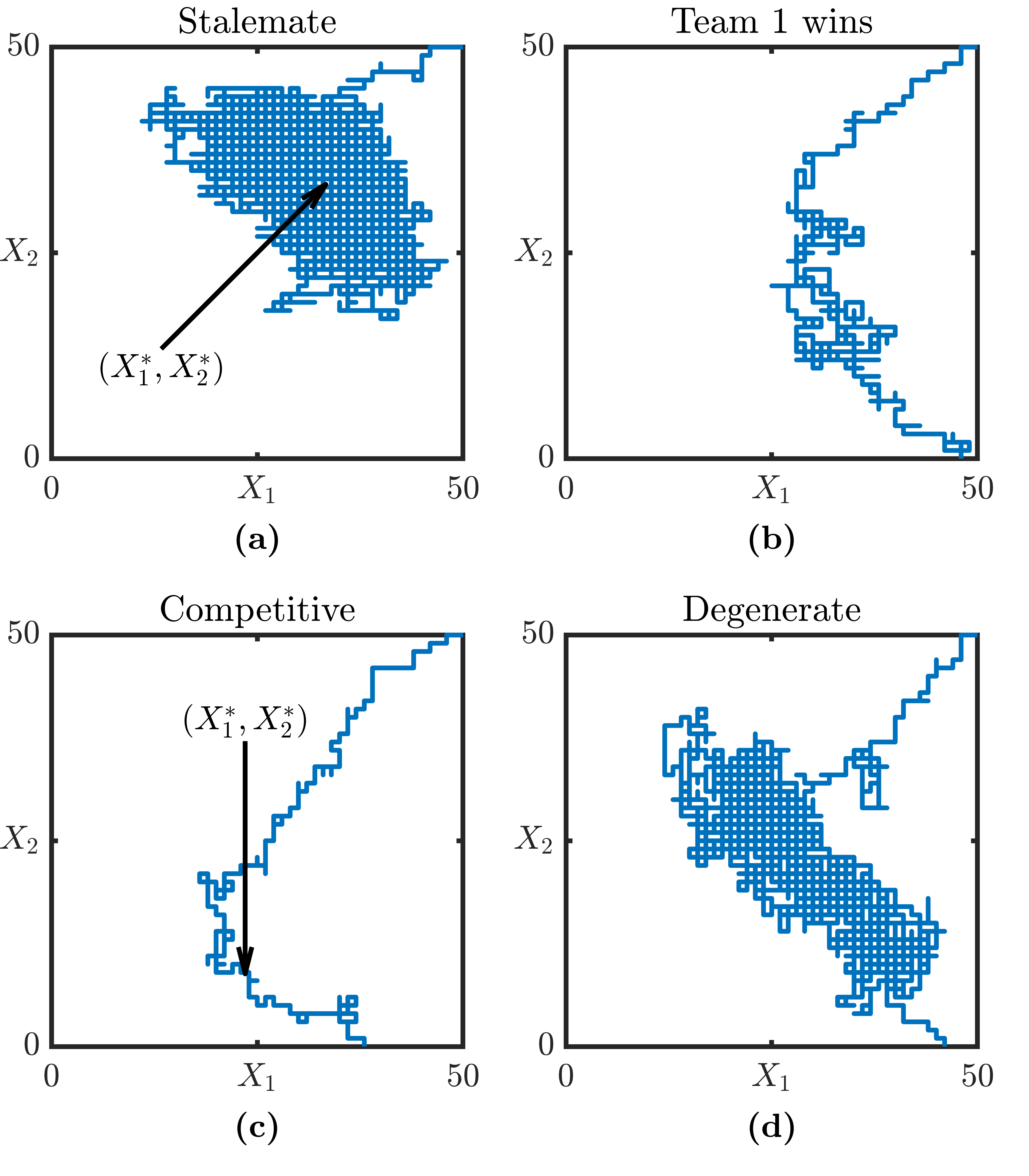}
    \caption{Simulations of games with the same constants as Fig.~\ref{fig:streams}. Trajectories $(X_1,X_2)$ have stochastic fluctuations on top of the deterministic flow of Fig~\ref{fig:streams}. The ``Stalemate'' regime (top left) results in long, back-and-forth games.}
    \label{fig:stoch_stream}
\end{figure}

\subsection{Stochastic games}

\begin{figure}[t]
    \centering
    \includegraphics[width=\linewidth]{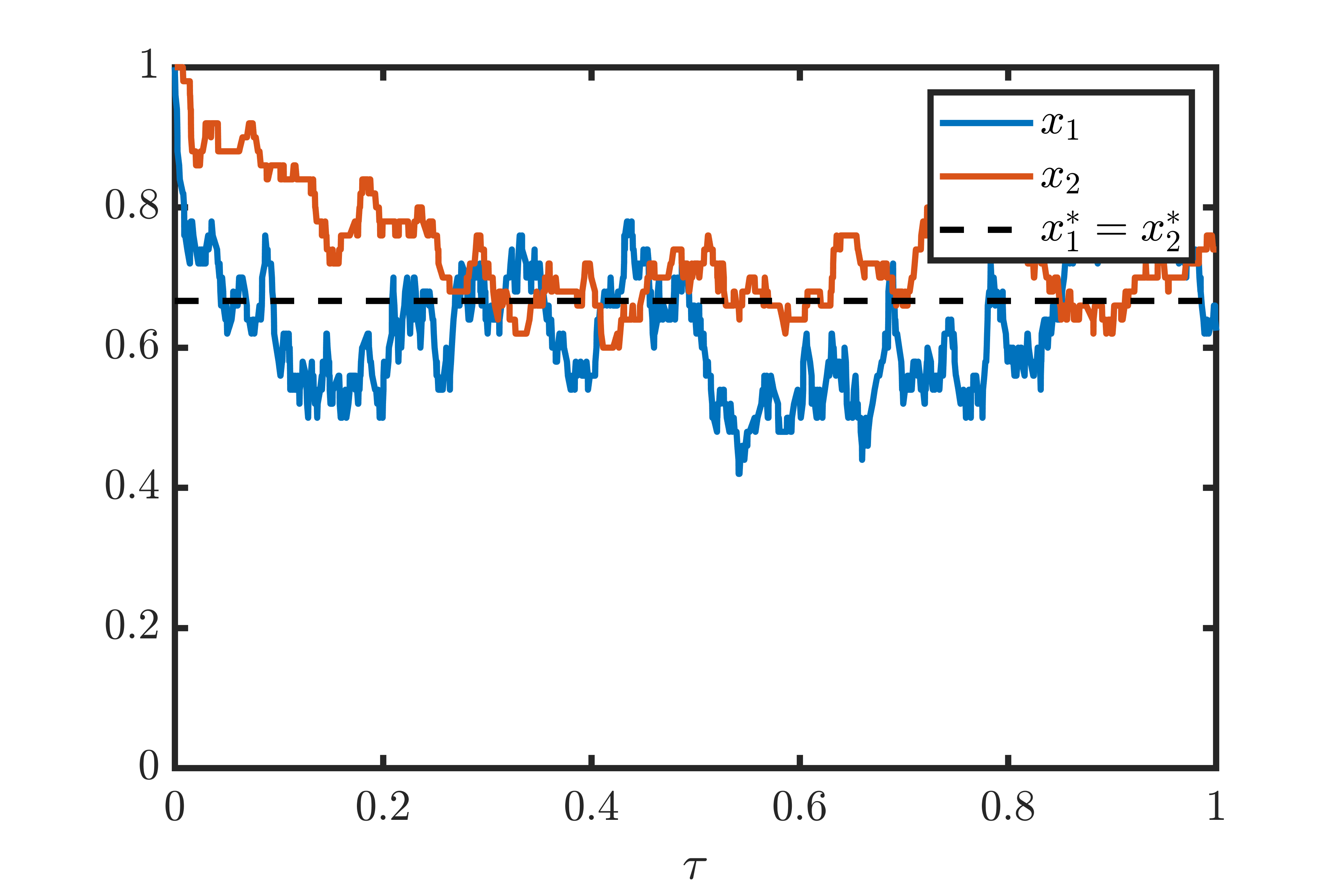}
    \caption{Fraction of players in Courts 1 and 2 (solid lines) versus dimensionless time $\tau$ for a stochastic game simulation with the same parameters as Fig~\ref{fig:streams} (top left), i.e., $c=1/2$, $a_1 = 1/4$, $a_2=3/4$, and $N=50$. In the ``Stalemate'' regime, the fraction of players fluctuates stochastically about the fixed point values $x_1^* = x_2^*$ (dashed line).}
    \label{fig:stalemate}
\end{figure}

In Figure~\ref{fig:stoch_stream} we show the evolution of four dodgeball games simulated as described above using the same parameters as in Fig.~\ref{fig:streams}. The plots show the trajectories of $(X_1,X_2)$ starting from initial conditions $(50,50)$. Note that, although the trajectories have significant fluctuations, they follow approximately the flow shown in Fig.~\ref{fig:streams}. In particular, for the parameters resulting in the stalemate scenario [i.e., a stable fixed point $(x_1^*,x_2^*) \in (0,1)\times(0,1)$] the number of players in Courts $1$ and $2$ fluctuates around $(N x_1^*,N x_2^*)$ (indicated with an arrow). In practice, these parameters result in extremely long games that continue until a random fluctuation is large enough to decrease $X_1$ or $X_2$ to zero. To further illustrate this, Fig.~\ref{fig:stalemate} shows $X_1(t)$ (blue) and $X_2(t)$ (orange) as a function of $t$ for the parameters in Fig.~\ref{fig:streams}(a). The evolution of this game resembles that of the games seen in Fig.~\ref{fig:game1}, which suggests that those games were in the Stalemate regime. In the degenerate case,  Fig.~\ref{fig:streams}(d), the game trajectory has large fluctuations around the line $X_1+X_2 = N$, which corresponds to the line of fixed points $x_1^* + x_2^* = 1$ of the deterministic system. We interpret this behavior as the trajectory diffusing under the effect of the fluctuations along the marginally stable line $X_1+X_2 = N$. Note that in the particular trajectory shown, Team 1 wins even after at some point in time they had only one player in Court 1. In Fig.~\ref{fig:stoch_stream}(c) the game eventually results in a victory by Team 1, even though the deterministic model predicts a victory by Team 2 [see~ Fig.~\ref{fig:streams}(c)], because stochastic fluctuations of the trajectory $(X_1,X_2)$ allow it to cross over to the basin of attraction of $(1,0)$.

As we see from these examples, the outcome of stochastic dodgeball games is determined both by the underlying deterministic flow and by the stochastic fluctuations of the $(X_1,X_2)$ trajectories. To account for this, we focus on how the probability $P$ of winning a game depends on the parameters. This probability can be calculated directly from the outcomes of a large number of simulated games (the algorithm for simulating games is presented in Appendix \ref{sec:appendixa}, but it is much more efficiently  calculated by using the properties of the underlying Markov process, as explained in Appendix \ref{sec:appendixb}. 
To illustrate how the probability of winning can be related to the deterministic results, we fix $c = 2/3$ and $a_2 = 3/4$, and calculate $P_1$ as a function of $a_1$.
\begin{figure}[t]
    \centering
    \includegraphics[width=\linewidth]{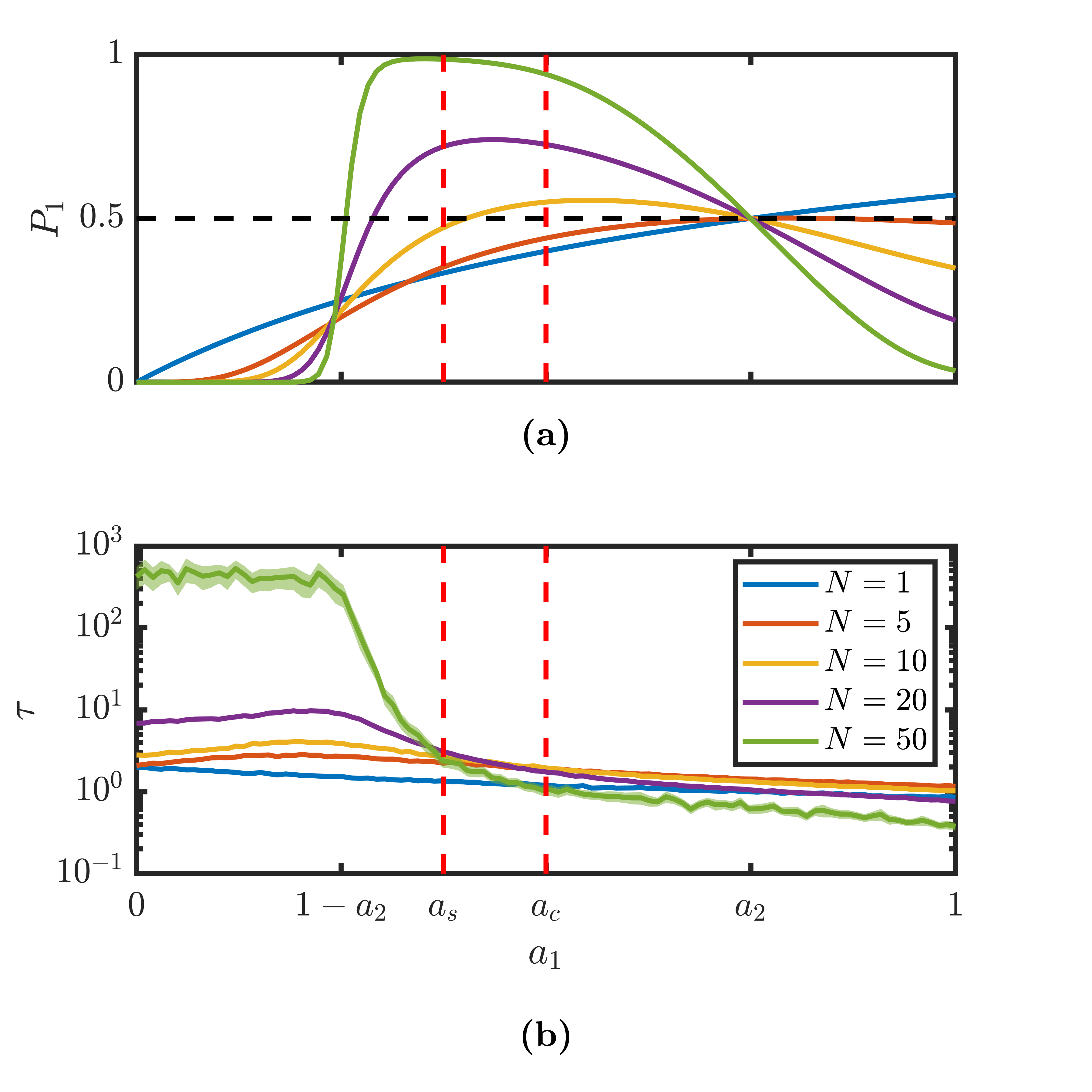}
    \caption{(a) Probability that Team 1 wins a game $P_1$ as a function of $a_1$ with $c=2/3$ and $a_2=3/4$ for $N = 1, 5, 10, 20,$ and $50$ (blue, orange, yellow, purple, and green solid lines, respectively). The dashed red lines mark bifurcations in the deterministic dynamics (see text), and the dashed horizontal line indicates $P_1 = 1/2$. The leftmost region corresponds to the ``Stalemate'' regime  leading to long games. The middle region represents ``Team 1 Wins'', which can be noted by the large values of $P_1$ for large values of $N$. The right region is the ``Competitive'' region in the deterministic model noted by mixed values of $P_1$ and quicker games. (b) Average duration of games (in dimensionless time $\tau$) with the same parameters as in the bottom panel. The duration of games in the ``Stalemate'' regime increases with $N$. The shaded area around the green curve represents $3$ standard deviations.}
    \label{fig:fix_sim}
\end{figure}
Fig~\ref{fig:fix_sim}(a) shows $P_1$ as a function of $a_1$ for $N = 1, 5, 10, 20,$ and $50$ (blue, orange, yellow, purple, and green solid lines, respectively). 
As $a_1$ increases from $0$ to $1$, different regimes of the deterministic model are traversed. For the parameters given let $a_s = (1-a_2)/c=3/8$ and $a_c = 1 - a_2 c=1/2$, which are shown as dashed red lines. For $0\leq a_1 <  a_s$, the system is in the ``Stalemate'' case, for  $a_s < a_1 <  a_c$ the system is the ``Team 1 Wins'' case, and for $a_c  < a_1 < 1$, it is in the ``Competitive'' case. Now we interpret how $P_1$ changes as $a_1$ is increased. For $a_1 < 1-a_2$, the fixed point $(x_1^*,x_2^*)$ is closer to $(0,1)$ than it is to $(1,0)$, and since victory is achieved by escaping the basin of attraction of the fixed point with random fluctuations, it is much more likely that this escape will occur to the nearest fixed point, in this case $(0,1)$. Therefore, $P_1\sim 0$ in this regime, and it is smaller for larger $N$ since fluctuations are smaller. For $1-a_2 < a_1 < a_s$, the game is still in the stalemate regime, but now $(x_1^*,x_2^*)$ is closer to $(1,0)$ and therefore $P_1 \sim 1$, and increases with $N$. For $a_s < a_1 < a_c$, the game is in the ``Team 1 Wins'' regime, and so $P_1$ approaches $1$ rapidly as $N$ increases. For $a_1 > a_c$, the game is in the ``Competitive'' regime, where the initial condition $(1,1)$ is in the basin of attraction of $(1,0)$ for $a_1 < a_2$ and in the basin of attraction of $(0,1)$ for $a_1> a_2$, which is reflected by the fact that $P_1 > 1/2$ for $a_1 < a_2$ and $P_1 < 1/2$ for $a_2 < a_1$. We note that for very small $N$ (e.g., $N = 1, 5$), the predictions of the deterministic theory break down. This can be understood in the limiting case $N=1$ (blue curve), where the probability of winning can be calculated explicitly as $P_1 = a_1/(a_1+a_2) = 4a_1/(4a_1 + 3)$.

According to our interpretation, victory in the ``Stalemate'' regime is achieved by escaping the basin of attraction of the underlying stable fixed point $(x_1^*,x_2^*)$ via fluctuations induced by the finite number of players. Since these fluctuations become less important as the number of players increases, one would expect that the average time $\tau$ to achieve victory would (i) be largest in the ``Stalemate'' regime, and (ii) increase with $N$. Fig~\ref{fig:fix_sim}(b) shows the average game duration $\tau$ as a function of $a_1$, calculated from direct simulation of 5000 stochastic games when $N < 50$ and $100$ games when $N=50$. Consistent with the interpretation above, $\tau$ is much longer in the ``Stalemate'' regime and increases with $N$ [we have found that $\tau$ scales exponentially with $N$ (not shown), as one would expect for an escape problem driven by finite size fluctuations]. Furthermore, it is maximum approximately when $(x_1^*,x_2^*)$ is equidistant to $(0,1)$ and $(1,0)$, i.e., when $a_1=1-a_2$ [see Fig~\ref{fig:fix_sim}(a)]. 

\begin{figure}
    \centering
    \includegraphics[width=\linewidth]{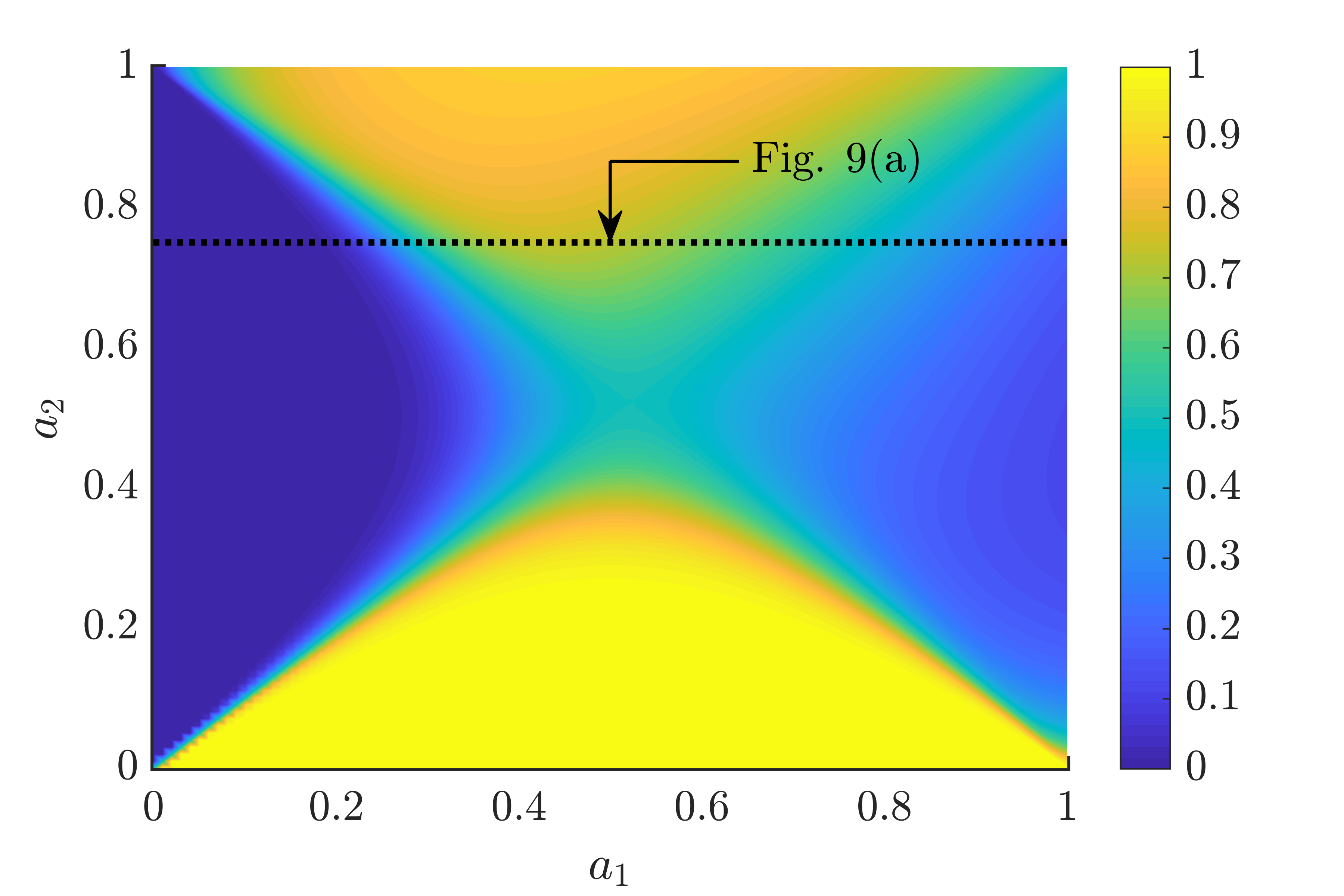}
    \caption{Probability that Team 1 wins  $P_1$ as a function of $a_1$ and $a_2$. The dashed line corresponds to the $N = 20$ curve in Fig.~\ref{fig:fix_sim}(a).}
    \label{fig:a1a2}
\end{figure}

To get a broader picture of how the choice of fixed strategies $a_1$, $a_2$ affects the probability of winning, we show in Fig.~\ref{fig:a1a2} the probability that Team 1 wins, $P_1$, as a function of $a_1$ and $a_2$, obtained numerically as described in Appendix~\ref{sec:appendixb} for $N=20$ and the same parameters of Fig.~\ref{fig:fix_point guide}(a). The curve for $N=20$ in Fig.~\ref{fig:fix_sim}(a) corresponds to the values shown in the dashed line.
There appears to be a saddle point approximately at $(a_1,a_2) \approx (1/2,1/2)$ corresponding to a Nash equilibrium, i.e., a set of strategies such that neither Team would benefit from a change of strategy if the other Team maintains their strategy. The issue of the appropriate definition and existence of Nash equilibria in finite-player stochastic games and their behavior as the number of players tends to infinity has been studied in the emerging area of {\it mean-field games} \cite{lasry2007mean,bensoussan2013mean}. We leave a more detailed study of Nash equilibria in dodgeball for future study.

\subsection{Heuristic Strategy} 

In the example treated in the previous Sections, the probability that a player in Team $i$ decides to throw a ball to an enemy player instead of rescuing a teammate from jail, $F_i(X_1,X_2)$ is fixed throughout the game at the value $a_i$. In reality, players may adjust this probability in order to optimize the probability of winning. In this Section we will develop a heuristic greedy strategy with the goal of trying to optimize victory. For this purpose, it is useful to define the quantities $H_i$ as
\begin{equation}
    \begin{array}{cc}
        H_1 = \frac{X_1}{X_1+X_2}, & H_2= \frac{X_2}{X_1+X_2}.
    \end{array} 
    \label{eqs:H}
\end{equation}
These quantities have the advantage that they are normalized between $0$ and $1$, with $H_i=0$ ($H_i=1$) corresponding to a loss (victory) by Team $i$. In addition, $H_i$ corresponds to the probability that team $i$ will throw a ball next, and therefore it is a good indicator of how much control team $i$ has. Therefore, it is reasonable for Team $i$ to apply a strategy to increase $H_i$. To develop such a strategy, we define $H_i$ and $H_i^{+}$ as the values of $H_i$ before and after  a ball is thrown. Similarly, we define $X_i$ and $X_i^+$ as the values of $X_i$ before and after a ball is thrown. For definiteness, we will present the strategy for Team $1$, and the strategy for Team $2$ will be similar. The basis of the strategy is to choose the value of $F_1(X_1,X_2)$ that maximizes the expected value of $H_1^{+}$, $\mathbb{E}[H_1^{+}]$. Since $F_1$ is the probability that the ball is thrown at enemy players, $p_e$ the probability that such a ball actually hits an enemy player, $1-F_1$ the probability that the ball is thrown at a teammate in jail, and $p_j$ the probability that such a ball is successful in rescuing a teammate, the expected value of $H_1^+$ is given by
\begin{multline}
    \mathbb{E}[H_1^{+}] = F_1\bigg[\frac{X_1}{X_1+X_2-1}p_e +\frac{X_1}{X_1+X_2}(1-p_e)\bigg]\\
    +(1-F_1)\bigg[\frac{X_1+1}{X_1+X_2+1}p_j + \frac{X_1}{X_1+X_2}(1-p_j)\bigg],
\end{multline}
Which can be rewritten as
\begin{equation}
    \mathbb{E}[H_1^{+}] = A + \frac{B}{X_1+X_2} F_1,
    \label{eq:linEx}
\end{equation}
where
\begin{align}
B = \bigg[\frac{X_1^t}{X_1^t+X_2^t-1}p_e - \frac{X_2^t}{X_1^t+X_2^t+1}p_j\bigg]
\end{align}
and $A$ is independent of $F_1$. 

Since Eq.~(\ref{eq:linEx}) is linear in $F_1$, it is maximized by choosing $F_1 = 1$ when $B > 0$ and $F_1 = 0$ when $B < 0$. Therefore, the choice of $F_1$ that maximizes the expected value of $H_1^{+}$, $F_1^*$, is 
\begin{equation}
    F_1^* = \begin{cases}
    1, & \frac{X_1}{X_1+X_2-1}p_e(X_2) \ge \frac{X_2}{X_1+X_2+1}p_j(N-X_1),\\
    0, & \text{otherwise}.
    \end{cases}
    \label{eq:disSol}
\end{equation}
When $X_1$, $X_2 \gg 1$, the strategy simplifies to
\begin{equation}
    F_1^* \approx \begin{cases}
    1, & X_1 p_e(X_2) \ge X_2 p_j(N-X_1),\\
    0, & \text{otherwise}.
    \end{cases}
    \label{eq:disSolApprox}
\end{equation}
We note that this can also be derived by maximizing $dH_1/dt$ by using Eqs.~(\ref{eq:simplified_a})-(\ref{eq:simplified_b}). Furthermore, for the case considered in Sections~\ref{sec:dyn} and \ref{sec:stoch}, where $p_e(X_i) = k_e X_i$ and $p_j(Y_i) = k_j Y_i$, the strategy reduces to 
\begin{equation}
    F_1^* = \begin{cases}
    1, & k_e X_1 \ge k_j(N-X_1),\\
    0, & \text{otherwise}.
    \end{cases}
    \label{eq:disSolEx}
\end{equation}
For example, when $k_e = k_j$ (i.e., the probability of success in hitting an enemy player is the same as the probability of  succeeding in rescuing a teammate from jail) the strategy for Team 1 consists in trying always to rescue teammates from Jail 1 when the majority of Team 1 player's are in Jail 1, and in trying to hit players from Team 2 when the majority of Team 1's players are in Court 1. Interestingly, in the limit $X_1$, $X_2 \gg 1$ the strategy for Team 1 is independent of $X_2$.

\begin{figure}[t]
    \centering
    \includegraphics[width=\linewidth]{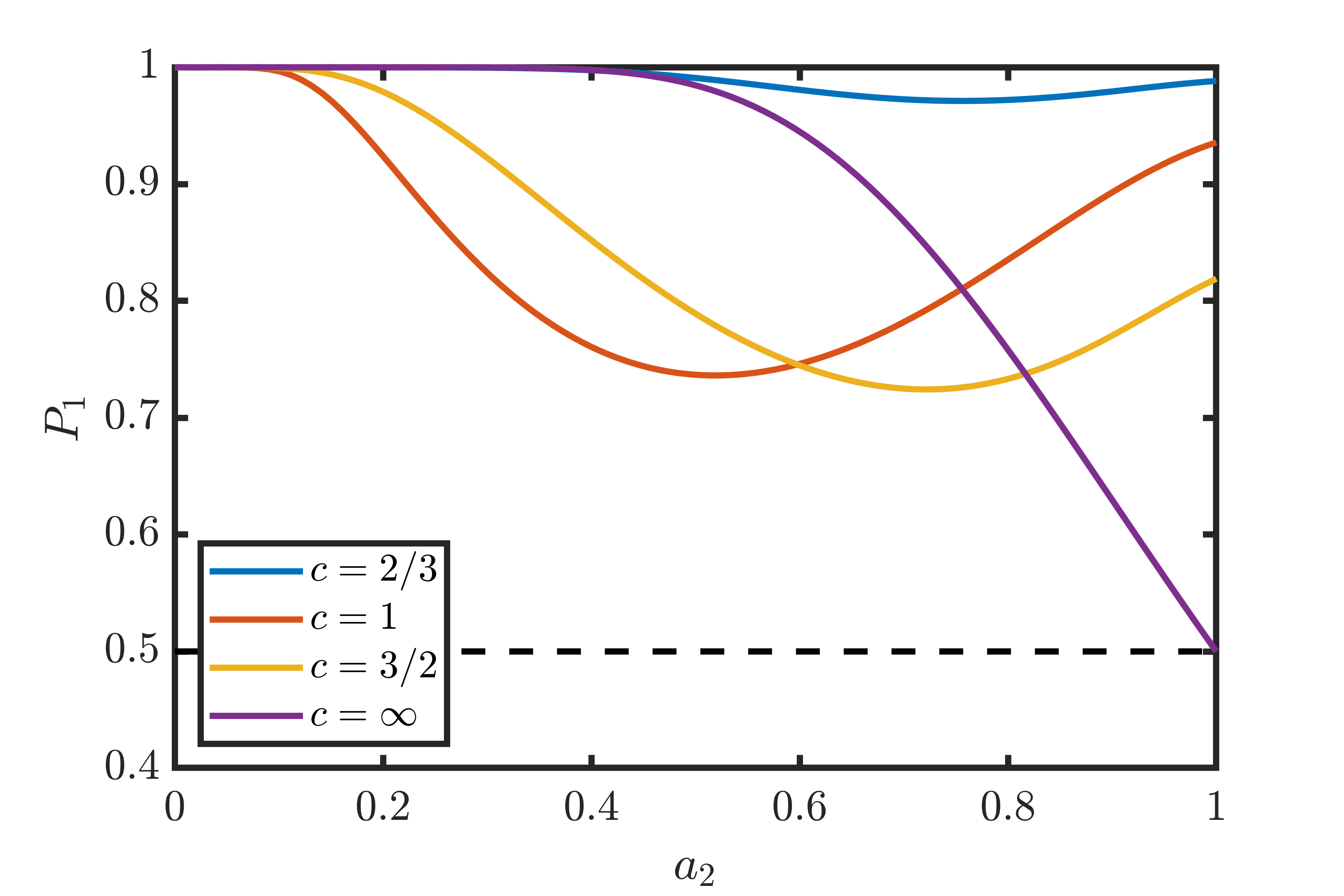}
    \caption{Probability of Team 1 winning with the heuristic strategy $F_1$ against a fixed strategy $a_2$. Number of players in each game is set to $N=20$. }
    \label{fig:strategy_plot}
\end{figure}

To validate the effectiveness of this strategy, we simulate dodgeball games in which Team 1 adopts the strategy $F_1(X_1,X_2) = F_1^*$ given by Eq.~(\ref{eq:disSol}) and Team 2 uses the fixed strategy $F_2(X_1,X_2) = a_2$. In Fig.~\ref{fig:strategy_plot} we plot the probability that Team 1 wins, $P_1$, as a function of $a_2$ for $c = 2/3, 1, 3/2,$ and $\infty$ (blue, orange, yellow, and purple solid lines, respectively). As the Figure shows, using the Strategy $F_1^*$ consistently results in a probability of winning higher than $1/2$. In general, the strategy $F_1^*$ does best when $c$ is small and $N$ is large. Note the probability of Team $1$ winning is $1/2$ only when $c=\infty$, i.e., the chance of saving a player in jail is $0$. In this case the strategy $a_2 = 1$ is clearly optimal.

\section{Conclusions}\label{sec:conclusions}
In this paper we presented a mathematical model of dodgeball, which we analyzed via an ODE-based compartmental model and numerical simulations of a stochastic agent-based model. These two complementary methods of analysis revealed a rich dynamical landscape. Depending on Teams' strategies, the dynamics and outcome of the game is determined by a combination of the stability of the fixed points of the underlying dynamical system and the stochastic fluctuations caused by the random behavior of individual players. 
Additionally, we derived a greedy strategy in the context of the stochastic model of dodgeball. While our strategy was shown to be effective against fixed strategies (i.e., $F_2 = a_2$), it isn't necessarily optimal. This suggests the future work of finding an optimal strategy as well as studying the topic of Nash equilibriums in the context of dodgeball.

   More data is needed to verify some of the predictions of the dodgeball model. While the time series from real games shown in Fig.~\ref{fig:game1} appear to be consistent with the Stalemate regime, a quantitative comparison would need estimation of the quantities $k_e$, $k_j$, $a_1$, and $a_2$. In principle, these probabilities could be estimated from recorded dodgeball games. Nevertheless, the continuous model of dodgeball is able to offer reasonable insights into the behavior of stochastic agent-based games with a realistic number of players.

    Our model and analysis relied on various assumptions and simplifications, and relaxing some of these assumptions could be a useful topic for future work as well. One significant assumption used is that a ball thrown at an enemy player will not be caught. However, it is possible for balls to be caught, and this causes the thrower to be sent to jail. The dodgeball model could be extended to include this situation. Who a player decides to target currently only depends on the number of remaining enemies in play and the number of people in jail, but this could be generalized to account for heterogeneous targeting probabilities. The last assumption that will be discussed here is that this model assumes uniform behavior of the players. Individual ability could be modeled by including an individual's ability to catch balls, hit an enemy target, and hit shots on jail. Finally, we assumed that players behave independently (which is a reasonable approximation in Elementary School games). Coordinated strategies such as those used in professional games are not considered here.

\acknowledgments
We thank James Meiss, Nicholas Landry, Daniel Larremore, and Max Ruth for their useful comments. We also thank Eisenhower Elementary for allowing us to use the data.

\appendix

\section{}

In this Appendix we provide details about the numerical simulation of the stochastic dodgeball games, and the numerical computation of winning probabilities $P_i$.

\subsection{Agent-based stochastic simulations}\label{sec:appendixa}

Here we describe the simulation of a single, stochastic agent-based dodgeball game. At $t = 0$, the game starts with $N$ players on each team, $X_1 = X_2 = N$. Since $\lambda$ is the rate at which players throw balls, and we assume that players throw balls independently of each other, the next ball throw in the game is exponentially distributed with rate
\begin{equation}
    r = (X_1 + X_2)\lambda.
\end{equation}
The probability that team $i$ throws a ball next (before the other team), which we denote $p_i$, is given by
\begin{align}
    p_1 = \frac{X_1}{X_1 + X_2},\\
    p_2 = \frac{X_2}{X_1 + X_2}.
    \label{eq:prob_choiceb}
\end{align}
The pseudo-code for simulating a game is below. Recall that $F_i(X_1,X_2)$ is the probability that team $i$ throws a ball towards the enemy instead of towards their jail, $p_e$ is the probability that a ball thrown towards the enemy hits a target, and $p_j$ is the probability that a ball thrown towards jail is successfully caught. In addition, we stop the simulation if the number of throws $k$ exceeds  $K_{\text{max}} = 50N^2$.
\begin{algorithm}[H]
\caption{Simulate dodgeball game}
\begin{algorithmic}
\STATE At $t = 0$, set $X_1 = X_2 = N$ and $k=0$.
\WHILE{($X_1 > 0$ and $X_2 > 0$) and $k \leq K_{\text{max}}$}
\STATE $k \leftarrow k + 1$\\
\STATE $t \leftarrow t\ +$ Exponential random variable with mean $1/r$.\\
\STATE Choose throwing team, $1$ or $2$, with probabilities $p_1$, $p_2$. Let the throwing team be $i$ and the other team be $j$.
\STATE Choose to throw ball at enemy or rescue from jail\\
\STATE with probabilities $F_i(X_1,X_2)$, $1-F_i(X_1,X_2)$.
\IF{Throw to enemy}
\STATE $X_j \leftarrow X_j -1$ with probability $p_e(X_j)$
\ENDIF
\IF{Throw to jail}
\STATE $X_i \leftarrow X_i + 1$ with probability $p_j(N-X_i)$
\ENDIF
\ENDWHILE
\end{algorithmic}
\end{algorithm}

\subsection{Calculation of winning probabilities $P_i$}\label{sec:appendixb}

Here we explain how the probability that team $i$ wins, $P_i$, is calculated for a given set of parameters.

First, we define as $\vec{v}_k$ the column vector whose entries are the probabilities that the game is in each of the $(N+1)^2$ possible states $(X_1,X_2)$ after the $k$-th ball is thrown.  Accordingly, $\vec{v}_0$ is the vector that represents the initial condition $(N,N)$. Then, we define $M$ as the $(N+1)^2\times (N+1)^2$ matrix of transition probabilities between these states. Because the game is a Markov process, we have 
\begin{equation}
    \vec{v}_{k+1} = M\vec{v}_k.
    \label{eq:iter_quick}
\end{equation}
Now we let $\vec{u}_i$ be a vector that is $1$ in each state in which Team $i$ wins and $0$ otherwise. Then,
\begin{equation}
    P_i = \lim_{k\rightarrow \infty} \vec{v}_k^T \vec{u}_i
    \label{eq:prob_lim}
\end{equation}
In practice, we stop the iteration when 
\begin{equation}
 |P_1+P_2 - 1|=|\vec{v}_k\cdot(\vec{u}_1+\vec{u}_2) - 1| < 10^{-4},
\end{equation}
or $k > K_{\text{max}} = 50N^2$. When the game is in stalemate, the expected length of games grows exponentially with $N$, and the calculation above becomes impractical for moderate values of $N$. In this case, we instead evolve the vector $\vec{v}_k$ in steps that are powers of two, as 
\begin{equation}
    \vec{v}_{2^j} =  M^{2^j} \vec{v}_0 =  (M^{2^{j-1}})^2 \vec{v}_0,
    \label{eq:iter_stalemate}
\end{equation}
In practice we stop this iteration when $j> J_\text{max}=256$. The iteration described by Eq.~\ref{eq:iter_stalemate} uses repeated non-sparse matrix multiplications, while Eq.~\ref{eq:iter_quick} uses faster sparse matrix-vector products. However, since games can be extremely long in the stalemate regime, the method described by Eq.~\ref{eq:iter_stalemate} is still faster in that regime. We choose the values $J_\text{max}$ and $K_{max}$ such that in practice Eq.~\ref{eq:iter_quick} and Eq.~\ref{eq:iter_stalemate} take similar amounts of time in the stalemate regime. 

\bibliography{bibfile}

\end{document}